\documentclass [prd,twocolumn,superscriptaddress,floatfix,showpacs,amsmath,letterpaper] {revtex4}

\usepackage {graphicx}
\usepackage {bm}
\usepackage {color}
\usepackage [bookmarks=true,bookmarksnumbered=true,colorlinks=false,pdfborder={0 0 0}] {hyperref}
\usepackage [final] {microtype}

\newcommand {\MM} [1] {\ensuremath{#1}}
\newcommand {\tsp} [1] {\ensuremath{\mskip #1\thinmuskip}}
\newcommand {\IT} [1] {\ensuremath{#1}}
\newcommand {\RM}  [1] {\ensuremath{\textrm{#1}}}
\newcommand {\SQRT}[1] {\ensuremath{\sqrt{#1}}}

\newcommand {\collision} [2] {\MM{#1 + #2}}
\newcommand {\SUB} [2] {\MM{#1\ensuremath{_{#2}}}}
\newcommand {\SUP} [2] {\MM{#1\ensuremath{^{#2}}}}

\newcommand {\DIVe}[2] {\MM{#1\tsp{-1}/\tsp{-0.3}#2}}

\newcommand {\momentum} {\IT{p}}

\newcommand {\transverse} {\IT{T}}

\newcommand {\pT} {\SUB{\momentum}{\transverse}}

\newcommand {\cspeed} {\IT{c}}

\newcommand {\pico} {\RM{p}}
\newcommand {\nano} {\RM{n}}

\newcommand {\milli} {\RM{m}}

\newcommand {\Giga} {\RM{G}}

\newcommand {\Volt} {\RM{V}}

\newcommand {\barn} {\RM{b}}
\newcommand {\mb} {\milli\barn}

\newcommand {\pb} {\pico\barn}
\newcommand {\nb} {\nano\barn}

\newcommand {\pbinv} {\SUP{\pb}{-\tsp{-0.4}1}}
\newcommand {\nbinv} {\SUP{\nb}{-\tsp{-0.4}1}}
\newcommand {\eV} {\tsp{0.1}\RM{e}\tsp{-0.3}\Volt}
\newcommand {\GeV} {\Giga\eV}
\newcommand {\GeVc} {\DIVe{\GeV}{\cspeed}}

\newcommand {\Tesla} {\RM{T}}

\newcommand {\rad} {\RM{rad}}

\newcommand {\unitns} [1] {\MM{#1}}
\newcommand {\unit} [1] {\MM{\tsp{1.5} #1}}

\newcommand {\PLMN} {\ensuremath{\pm}}

\newcommand {\PI} {\ensuremath{\pi}}

\newcommand {\piplus} {\SUP{\PI}{+}}
\newcommand {\piminus} {\SUP{\PI}{-}}
\newcommand {\piplusminus} {\SUP{\PI}{\PLMN}}
\newcommand {\pizero} {\SUP{\PI}{0}}
\newcommand {\svar} {\IT{s}}

\newcommand {\sqrts} {\SQRT{\tsp{-0.5}\svar\tsp{0.5}}}

\newcommand {\momentumthree}{\textbf{p}}

\newcommand {\proton} {\IT{p}}

\newcommand {\protonproton} {\collision{\proton}{\proton}}

\newcommand {\meanz} {\ensuremath{\langle}\IT{z}\ensuremath{\rangle}}

\newcommand {\program} [1] {\textsc{#1}}

\newcommand {\PYTHIA} {\program{pythia}}
\newcommand {\GEANT} {\program{geant}}

\newcommand {\capsword} [1] {\textsc{#1}}

\newcommand {\NLO} {\capsword{NLO}}
\newcommand {\QCD} {\capsword{QCD}}
\newcommand {\pQCD} {p\capsword{QCD}}
\newcommand {\DIS} {\capsword{DIS}}

\newcommand {\BNL} {\capsword{BNL}}
\newcommand {\RHIC} {\capsword{RHIC}}
\newcommand {\RCF} {\capsword{RCF}}
\newcommand {\NERSC} {\capsword{NERSC}}
\newcommand {\LBNL} {\capsword{LBNL}}

\newcommand {\PHENIX} {\capsword{PHENIX}}
\newcommand {\STAR} {\capsword{STAR}}

\newcommand {\CDF} {\capsword{CDF}}

\newcommand {\CTEQsixM} {\capsword{CTEQ6M}}
\newcommand {\DSS} {\capsword{DSS}}
\newcommand {\DSSV} {\capsword{DSSV}}
\newcommand {\GRSV} {\capsword{GRSV}}
\newcommand {\GSC} {\capsword{GS-C}}
\newcommand {\IR} {\capsword{IR}}
\newcommand {\CNI} {\capsword{CNI}}
\newcommand {\BEMC} {\capsword{BEMC}}
\newcommand {\BSMD} {\capsword{BSMD}}
\newcommand {\BBC} {\capsword{BBC}}
\newcommand {\TPC} {\capsword{TPC}}

\newcommand {\MB} {\capsword{MB}}
\newcommand {\HT} {\capsword{HT}}

\hypersetup {
pdfauthor = {STAR Collaboration},
pdftitle = {Longitudinal double-spin asymmetry and cross section for inclusive neutral pion production at midrapidity in polarized proton collisions at sqrt(s) = 200 GeV},
pdfkeywords = {polarized proton collisions,high transverse momentum,neutral pion production,longitudinal double-spin asymmetry}
}

\begin {document}

\title {Longitudinal double-spin asymmetry and cross section for inclusive neutral pion production at midrapidity in polarized proton collisions at $\protect\bm{\sqrts = 200\unit{\GeV}}$
\enlargethispage{2\baselineskip}
}

\affiliation{Argonne National Laboratory, Argonne, Illinois 60439, USA}
\affiliation{University of Birmingham, Birmingham, United Kingdom}
\affiliation{Brookhaven National Laboratory, Upton, New York 11973, USA}
\affiliation{University of California, Berkeley, California 94720, USA}
\affiliation{University of California, Davis, California 95616, USA}
\affiliation{University of California, Los Angeles, California 90095, USA}
\affiliation{Universidade Estadual de Campinas, Sao Paulo, Brazil}
\affiliation{University of Illinois at Chicago, Chicago, Illinois 60607, USA}
\affiliation{Creighton University, Omaha, Nebraska 68178, USA}
\affiliation{Czech Technical University in Prague, FNSPE, Prague, 115 19, Czech Republic}
\affiliation{Nuclear Physics Institute AS CR, 250 68 \v{R}e\v{z}/Prague, Czech Republic}
\affiliation{University of Frankfurt, Frankfurt, Germany}
\affiliation{Institute of Physics, Bhubaneswar 751005, India}
\affiliation{Indian Institute of Technology, Mumbai, India}
\affiliation{Indiana University, Bloomington, Indiana 47408, USA}
\affiliation{University of Jammu, Jammu 180001, India}
\affiliation{Joint Institute for Nuclear Research, Dubna, 141 980, Russia}
\affiliation{Kent State University, Kent, Ohio 44242, USA}
\affiliation{University of Kentucky, Lexington, Kentucky, 40506-0055, USA}
\affiliation{Institute of Modern Physics, Lanzhou, China}
\affiliation{Lawrence Berkeley National Laboratory, Berkeley, California 94720, USA}
\affiliation{Massachusetts Institute of Technology, Cambridge, MA 02139-4307, USA}
\affiliation{Max-Planck-Institut f\"ur Physik, Munich, Germany}
\affiliation{Michigan State University, East Lansing, Michigan 48824, USA}
\affiliation{Moscow Engineering Physics Institute, Moscow Russia}
\affiliation{City College of New York, New York City, New York 10031, USA}
\affiliation{NIKHEF and Utrecht University, Amsterdam, The Netherlands}
\affiliation{Ohio State University, Columbus, Ohio 43210, USA}
\affiliation{Old Dominion University, Norfolk, VA, 23529, USA}
\affiliation{Panjab University, Chandigarh 160014, India}
\affiliation{Pennsylvania State University, University Park, Pennsylvania 16802, USA}
\affiliation{Institute of High Energy Physics, Protvino, Russia}
\affiliation{Purdue University, West Lafayette, Indiana 47907, USA}
\affiliation{Pusan National University, Pusan, Republic of Korea}
\affiliation{University of Rajasthan, Jaipur 302004, India}
\affiliation{Rice University, Houston, Texas 77251, USA}
\affiliation{Universidade de Sao Paulo, Sao Paulo, Brazil}
\affiliation{University of Science \& Technology of China, Hefei 230026, China}
\affiliation{Shandong University, Jinan, Shandong 250100, China}
\affiliation{Shanghai Institute of Applied Physics, Shanghai 201800, China}
\affiliation{SUBATECH, Nantes, France}
\affiliation{Texas A\&M University, College Station, Texas 77843, USA}
\affiliation{University of Texas, Austin, Texas 78712, USA}
\affiliation{Tsinghua University, Beijing 100084, China}
\affiliation{United States Naval Academy, Annapolis, MD 21402, USA}
\affiliation{Valparaiso University, Valparaiso, Indiana 46383, USA}
\affiliation{Variable Energy Cyclotron Centre, Kolkata 700064, India}
\affiliation{Warsaw University of Technology, Warsaw, Poland}
\affiliation{University of Washington, Seattle, Washington 98195, USA}
\affiliation{Wayne State University, Detroit, Michigan 48201, USA}
\affiliation{Institute of Particle Physics, CCNU (HZNU), Wuhan 430079, China}
\affiliation{Yale University, New Haven, Connecticut 06520, USA}
\affiliation{University of Zagreb, Zagreb, HR-10002, Croatia}

\author{B.~I.~Abelev}\affiliation{University of Illinois at Chicago, Chicago, Illinois 60607, USA}
\author{M.~M.~Aggarwal}\affiliation{Panjab University, Chandigarh 160014, India}
\author{Z.~Ahammed}\affiliation{Variable Energy Cyclotron Centre, Kolkata 700064, India}
\author{A.~V.~Alakhverdyants}\affiliation{Joint Institute for Nuclear Research, Dubna, 141 980, Russia}
\author{B.~D.~Anderson}\affiliation{Kent State University, Kent, Ohio 44242, USA}
\author{D.~Arkhipkin}\affiliation{Brookhaven National Laboratory, Upton, New York 11973, USA}
\author{G.~S.~Averichev}\affiliation{Joint Institute for Nuclear Research, Dubna, 141 980, Russia}
\author{J.~Balewski}\affiliation{Massachusetts Institute of Technology, Cambridge, MA 02139-4307, USA}
\author{O.~Barannikova}\affiliation{University of Illinois at Chicago, Chicago, Illinois 60607, USA}
\author{L.~S.~Barnby}\affiliation{University of Birmingham, Birmingham, United Kingdom}
% ...

\collaboration{STAR Collaboration}\noaffiliation

\date {November 14, 2009}

\begin {abstract}
We report a measurement of the longitudinal double-spin asymmetry $A_{LL}$ and the differential cross section for inclusive \pizero\ production at midrapidity in polarized proton collisions at $\sqrts = 200\unit{\GeV}$\@. 
The cross section was measured over a transverse momentum range of $1 < \pT < 17\unit{\GeVc}$ and found to be in good agreement with a next-to-leading order perturbative \QCD\ calculation. 
The longitudinal double-spin asymmetry was measured in the range of $3.7 < \pT < 11\unit{\GeVc}$ and excludes a maximal positive gluon polarization in the proton. 
The mean transverse momentum fraction of \pizero's in their parent jets was found to be around $0.7$ for electromagnetically triggered events.
\end {abstract}


\maketitle

\enlargethispage{2\baselineskip}
The spin structure of the nucleon is one of the fundamental and unresolved questions in Quantum Chromodynamics (\QCD)\@. 
Deep-inelastic scattering (\DIS) experiments studying polarized leptons scattered off polarized nuclei have found the quark and anti-quark spin contributions 
to the overall spin of the nucleon to be small, at the level of $25\%$~\cite {Ashman:1989ig, Filippone:2001ux}, leading to increased 
interest in the spin contribution from gluons. \DIS\ experiments have placed coarse constraints on the polarized gluon distribution function 
$\Delta g (x)$, based on the scale dependence of polarized structure functions~\cite {Adeva:1998vw, Anthony:2000fn} and on recent semi-inclusive 
data~\cite {Airapetian:1999ib, Adeva:2004dh, Ageev:2005pq}\@. 
Measurements using collisions of longitudinally polarized protons are attractive because they provide sensitivity 
to the polarized gluon spin distribution at leading order through quark--gluon and gluon--gluon scattering contributions to the \mbox{cross section}.

The sensitivity of inclusive hadron and jet production to the underlying gluon polarization in high-energy polarized proton collisions 
has been discussed in detail in Refs.~\cite {Jager:2002xm,Jager:2004jh}\@. 
The theoretical framework in the context of next-to-leading order 
perturbative \QCD\ (\NLO\ \pQCD) calculations is very well developed to constrain $\Delta g(x)$\@.
The first global analysis of semi-inclusive and inclusive
\DIS\ data, as well as results obtained by the \PHENIX~\cite {Adare:2007dg} and \STAR~\cite {Abelev:2007vt} experiments, placed a strong constraint on $\Delta g(x)$ 
in the gluon momentum-fraction range of $0.05 < x < 0.2$, and suggested that the gluon spin contribution is not large in that range~\cite {deFlorian:2008mr}\@. 
This conclusion was driven primarily by data on inclusive hadron and jet production in polarized proton collisions at $\sqrts = 200\unit{\GeV}$ at \RHIC\@.

In this paper, we report on the measurement of the cross section and the longitudinal double-spin asymmetry $A_{LL}$ 
for inclusive \pizero\ production at midrapidity in polarized proton collisions at $\sqrts = 200\unit{\GeV}$ by 
the \STAR\ experiment~\cite {Ackermann:2002ad} at \RHIC\@. 
The cross section is compared to a \NLO\ \pQCD\ calculation and the observed agreement
provides an important basis to apply \pQCD\ for the interpretation of $A_{LL}$\@.
The asymmetry is defined as
\begin {equation}
\label {all_def}
A_{LL} \equiv \frac{\sigma^{++} - \sigma^{+-}}{\sigma^{++} + \sigma^{+-}},
\end {equation}
where $\sigma^{++}$ and $\sigma^{+-}$ are the inclusive \pizero\ cross sections for 
equal ($++$) and opposite ($+-$) beam helicity configurations.
The measured longitudinal double-spin asymmetry probes a gluon momentum fraction of approximately $0.03 < x < 0.3$,
and is compared to \NLO\ \pQCD\ calculations.
In addition, we present the mean transverse momentum fraction of \pizero's in electromagnetically triggered jets. 
This measurement allows one to relate the spin asymmetry measurements performed with inclusive \pizero's to those using reconstructed jets. 
It may also help to constrain fragmentation models. 

The data for the analyses presented here were collected at \STAR\ in 2005
using stored polarized $100\unit{\GeV}$ proton beams with an average luminosity of $6 \times 10^{30}\unit{\mathrm{cm}^{-2}\,\mathrm{s}^{-1}}$\tsp{-0.5}. 
Longitudinal polarization of proton beams in the \STAR\ interaction region (\IR) was achieved by spin rotator magnets upstream and 
downstream of the \IR\ that changed the proton spin orientation from its stable vertical direction to 
longitudinal~\cite {Alekseev:2003sk}\@. The helicities were alternated between successive proton bunches in 
one beam and pairs of successive proton bunches in the other beam. 
This allowed us to obtain all four helicity combinations of the colliding bunch pairs at the \STAR\ \IR\ in quick succession\@. 
Additional reduction of systematic uncertainties was achieved by periodically changing the
helicity patterns of the stored beams. 
The polarization of each beam was measured several times per fill 
using Coulomb--Nuclear Interference (\CNI) proton--carbon polarimeters~\cite {Nakagawa:2007zza}, 
which were calibrated using a polarized hydrogen gas-jet target~\cite {Makdisi:2007zz}\@. 
The average \RHIC\ beam polarizations in the 2005 run were $P_1 = 52 \pm 3\%$ and $P_2 = 48 \pm 3\%$\@.
Non-longitudinal beam polarization components were continuously monitored with local polarimeters at \STAR~\cite {Kiryluk:2005gg} and 
were found to be no larger than $9\%$ in absolute magnitude. 
 
The principal \STAR\ detector subsystems for the measurements presented here were the Barrel Electromagnetic Calorimeter 
(\BEMC)~\cite {Beddo:2002zx} and the Beam--Beam Counters (\BBC)~\cite {Kiryluk:2005gg}\@. 
In addition, the Time Projection Chamber (\TPC)~\cite {Anderson:2003ur} was used for vertexing, 
for measuring the charged component in the reconstructed jets,
and as a charged particle veto for the \pizero\ reconstruction. 
The \BEMC\ is a lead--scintillator sampling calorimeter with a granularity of 
$\Delta \eta \times \Delta \varphi = 0.05 \times 0.05\unit{\rad}$, where one such cell is referred to as a tower.  
It contains a shower maximum detector (\BSMD) that consists of two layers of wire proportional counters with cathode strip readout, one in 
the azimuthal direction and one in the longitudinal direction, at a depth of about $5$ radiation lengths in each calorimeter module, 
providing a segmentation of $0.007 \times 0.007\unit{\rad}$\@.
For the 2005 running period, half of the \BEMC\ was instrumented and operational, providing $2\pi$ azimuthal coverage for $0 < \eta < 1$\@.
The \BBC{}s are composed of segmented scintillator rings, covering $3.3 < \vert \eta \vert < 5.0$ on both sides of the \IR\@. 
The \BBC{}s were used to trigger on collisions, to measure the helicity-dependent relative luminosities, and to serve as local polarimeters. 
The \TPC\ provided charged particle tracking inside a $0.5\unit{\Tesla}$ solenoidal magnetic field over the full range of azimuthal angles for $\vert \eta \vert < 1.3$\@. 

Proton--proton collisions in the \STAR\ detector were identified by a minimum bias trigger (\MB), defined as a coincidence of 
hits in both \BBC{}s\@. The cross section for this trigger was $\sigma_{\text{\BBC}} = 26.1 \pm 0.2\,\text{(stat)} \pm 1.8\,\text{(syst)}\unit{\mb}$, 
corresponding to $87 \pm 8\%$ of the non-singly diffractive \protonproton\ cross section at $\sqrts = 200\unit{\GeV}$~\cite {Adams:2003kv}\@. 
Rare hard scattering events were selected by two high-tower triggers, \HT1 and \HT2,
that required a transverse energy deposition in a single \BEMC\ tower
above thresholds of $2.6$ and $3.5\unit{\GeV}$\tsp{-0.5}, respectively,
in addition to satisfying the \MB\ condition.

A data sample with an integrated luminosity of $\mathcal{L} = 0.17\unit{\nbinv}$ for \MB, 
$0.16\unit{\pbinv}$ for \HT1, and $0.66\unit{\pbinv}$ for \HT2 triggers 
was analyzed for the inclusive cross section measurement.
Data with an integrated luminosity of $0.4\,(2.0)\unit{\pbinv}$ of \HT1$\,$(\HT2) triggers were used for the $A_{LL}$ determination.
The event selection criteria for the asymmetry analysis were identical to those used in a previously published jet measurement~\cite {Abelev:2007vt}\@.
About $22\%$ of \HT1/\HT2 triggered events also entered the jet $A_{LL}$ measurement~\cite {Abelev:2007vt},
but represented a negligible fraction of the much larger inclusive jet data set.
Therefore, the statistical correlation of the present \pizero\ and jet $A_{LL}$ measurements is negligible.

\enlargethispage{2\baselineskip}
Neutral pions were reconstructed in the decay channel $\pizero\rightarrow\gamma\gamma$
in an invariant mass analysis of pairs of neutral \BEMC\ clusters, i.e., those that did not have a \TPC\ track pointing to them,
with a cut on the two-particle energy asymmetry of $|E_1-E_2|/(E_1+E_2) \le 0.7$\@. 
The tower granularity was insufficient to resolve cluster pairs in \HT1/\HT2 data 
because of the small opening angle between daughter photons of pions that satisfied these triggers.
Therefore, the \BSMD\ clusters were used to determine the photon coordinates in those data.
A fiducial volume cut on the detector pseudorapidity of $0.1 < \eta < 0.9$ was imposed.
The reconstructed value of the pion pseudorapidity with respect to the vertex position was required to fall in the range $0 < \eta < 1$\@.
The \pizero\ yield was extracted in $\pT$ bins by integrating the background-subtracted invariant mass
distribution in a $\pT$-dependent window around the $\pizero$ peak that corresponded to an approximately $\pm\,3\,\sigma$ range. 
The combinatorial background was determined using the event mixing method with a jet alignment correction~\cite {ref_grebenyuk_thesis,pi0_pp_dau}\@. 

\begin {figure}
\includegraphics {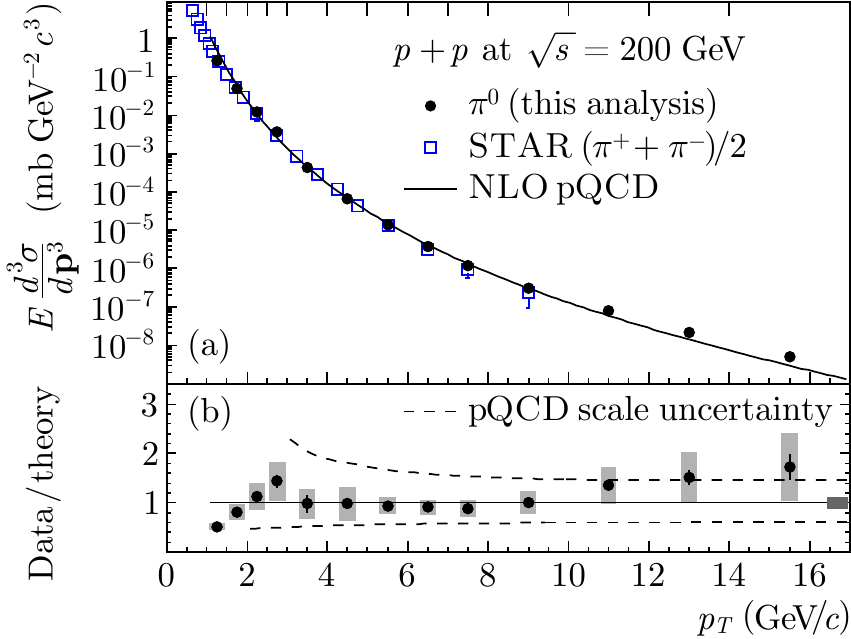}
\caption {\label {fig:XSect}(color online) (a) Cross section for inclusive \pizero\ production 
at midrapidity in \protonproton\ collisions at $\sqrts = 200\unit{\GeV}$\tsp{-0.5}, 
compared to a \NLO\ \pQCD\ calculation~\protect\cite {Jager:2002xm} based 
on the \DSS\ set of fragmentation functions~\protect\cite {deFlorian:2007aj}, 
and to the \STAR\ $\piplusminus$ measurement~\protect\cite {Adams:2006nd}\@. 
(b) The ratio of measured cross section and the \NLO\ \pQCD\ calculation.
The scale uncertainty is indicated by the dashed curves ($\mu = 2\pT$, $\pT/2$)\@. 
The error bars are statistical and shaded bands are $\pT$-correlated systematic uncertainties.
The normalization uncertainty is indicated by a shaded band around unity on the right-hand side.}
\end {figure}

The cross section for \pizero\ production is given by
\begin {equation}
E \frac{d^{\tsp{0.5}3} \sigma}{d\tsp{0.5}\momentumthree^3} = \frac{1}{2\pi \pT\,\Delta \pT \Delta \eta} \, c \, \frac{N}{\mathcal{L}},
\end {equation}
where $\Delta\pT$ and $\Delta \eta$ are the bin widths in $\pT$ and pseudorapidity, 
$N$ is the \pizero\ yield in a bin, and $c$ is an overall correction factor that 
accounts for acceptance, reconstruction, and trigger efficiency in that bin,
which was determined 
using a Monte Carlo simulation of \pizero's passed through
the \GEANT~\cite {Brun:1978fy} model of the \STAR\ detector.
Figure~\ref {fig:XSect} 
shows the differential cross section for inclusive \pizero\ production.
This analysis covered the pion transverse momentum range of $1 < \pT < 17\unit{\GeVc}$, 
and data points were scaled to the bin centers using local exponential fits around each bin.
The cross sections up to $4\unit{\GeVc}$ were measured using \MB\ triggered events; above $4\,$($7$)$\unit{\GeVc}$ the entries were obtained from \HT1$\,$(\HT2) triggers. 
The different trigger samples agreed within errors. 

The dominant systematic uncertainty ($25\%$ on average) of the measured cross section was due to 
a $5\%$ uncertainty in the global energy scale of the \BEMC\@. 
The other systematic uncertainties were related to yield extraction ($7\%$), 
reconstruction efficiency ($6\%$),
and relative normalization of \HT1/\HT2 and \MB\ triggers ($5\%$)\@.
An additional uncertainty due to the limited quality of the electromagnetic shower simulation at low photon energies in our \GEANT\ model was assigned to the cross section obtained from \HT1/\HT2 data 
[$15$($2$)$\unitns{\%}$ at $\pT = 4$($7$)$\unit{\GeVc}$]\@.

In Fig.~\ref {fig:XSect}, the measured cross section is compared to a \NLO\ \pQCD\ calculation~\cite {Jager:2002xm} performed using the \CTEQsixM\ set of unpolarized parton 
distribution functions~\cite {Pumplin:2002vw} and the \DSS\ set of fragmentation functions~\cite {deFlorian:2007aj}\@. 
In this calculation, the factorization and renormalization scales were identified 
with $\pT$ (solid curve), and were varied by a factor of two to estimate the impact of scale uncertainties (dashed curves)\@. 
The \DSS\ analysis included recent measurements of \pizero\ production at midrapidity by \PHENIX~\cite {Adare:2007dg} and at forward rapidity 
by \STAR~\cite {Adams:2006uz}\@.
The \NLO\ \pQCD\ calculation shows, within errors, good agreement with our data in the fragmentation region $\pT > 2\unit{\GeVc}$\@. 
We also compare the cross section for $\pizero$ production to the \STAR\ $\piplusminus$ measurement~\cite {Adams:2006nd}\@.
The \pizero\ and $(\piplus + \piminus)/2$ cross sections are expected to be equal, 
and the two \STAR\ measurements agree within statistical errors, in spite of using independent detector sub-systems.

The transverse momentum fraction carried by a high-$\pT$ \pizero\ in its parent jet, $z = \pT(\pizero)/\pT(\rm{jet})$, 
was investigated by associating pions with jets found in the same event~\cite {Abelev:2006uq}\@. 
The \pizero\ sample, defined by the invariant mass window, contained $\approx$\tsp{0.5}$8\unitns{\%}$ of combinatorial background.
An association was made if the pion was within a cone of radius $R = \sqrt{(\Delta\eta)^2 + (\Delta\varphi)^2} = 0.4$ around the jet axis. 
The analysis was restricted to $0.4 < \eta < 0.6$ in the jet pseudorapidity, 
so that the reconstructed jets were fully contained in the \BEMC\ acceptance. 
The transverse momentum of the jet was required to exceed $5\unit{\GeVc}$\@. 
The jet was required to have a neutral energy fraction less than $0.95$, in order to minimize contributions from beam background to the reconstructed jet sample.

\begin {figure}
\includegraphics {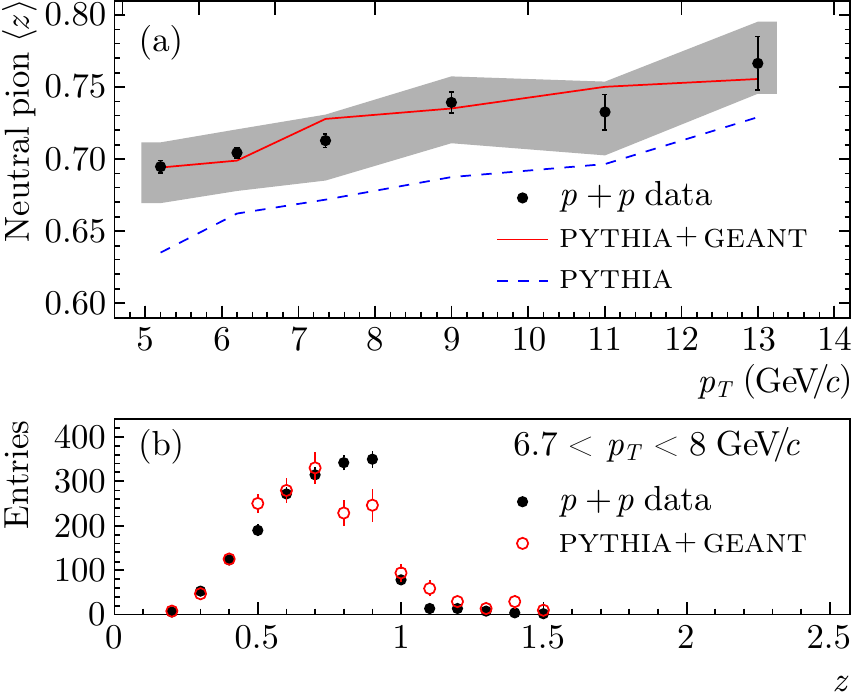}
\caption {\label {fig:MeanZ}(color online) (a) Mean transverse momentum fraction of \pizero's in their associated jets, 
as a function of pion $\pT$, for electromagnetically triggered events. 
Systematic errors are shown by the shaded band around the data points. 
The curves are results from simulations with the \PYTHIA\ event generator. 
The solid curve includes detector effects simulated by \GEANT, while the dashed curve uses jet finding at the \PYTHIA\ particle level. 
(b) The distribution of $z$ for one \pT\ bin, compared to \PYTHIA\ with a full detector response simulation.}
\end {figure}

Figure~\ref {fig:MeanZ}(a)
shows the mean value of $z$ as a function of pion $\pT$, combined for \HT1 and \HT2 triggers.
The data points are plotted at the bin centers in pion $\pT$\@.
The results were not corrected for detector effects, such as acceptance, efficiency, or resolution of the jet reconstruction. 
The systematic error band shown includes contributions from the uncertainty of the jet energy scale, the influence of the cut on minimum jet $\pT$, 
the contribution of events with $z > 1$, and a variation of other analysis cuts. 

The \meanz\ of \pizero's in electromagnetically triggered jets was found to be around $0.7$ and to rise slightly with pion $\pT$, 
consistent with measurements of leading charged hadrons in jets in fixed-target experiments~\cite {Boca:1990rh}\@. 
The results also compare well to recent theoretical calculations for charged pions \cite{deFlorian:2009fw}, 
considering the increase of the measured pion momentum fraction due to energy not reconstructed in the jet.  
The expectations from a \PYTHIA-based (version \capsword{6.205}~\cite {Sjostrand:2001yu} with `\CDF\ Tune \capsword{A}' settings~\cite {Field:2005sa}) Monte Carlo simulation are also shown. 
The \meanz\ measured in jets found on the \PYTHIA\ particle level, i.e., without any detector effects, 
is lower than in the data due to resolution effects and losses in the jet reconstruction, indicating the influence of the detector on the measurement. 
Results from a \GEANT-based \STAR\ detector simulation show good agreement with the data, 
demonstrating the reliability of the simulation framework used in the present analysis. 

\enlargethispage{\baselineskip}
Figure~\ref {fig:MeanZ}(b) shows the distribution of $z$ for one of the bins in pion \pT\ in 
comparison to \PYTHIA\ with a \GEANT-based detector simulation. 
To maximize the statistics in the simulation, the generator-level \pizero's were used without requiring an explicit reconstruction. 
This led to a softening of the falling edge of the distribution at high $z$ in simulations, 
since a full \GEANT\ simulation was used for the containing jets, but did not affect the mean of the distribution. 
A small fraction of the events had $z > 1$, apparently corresponding to pions that carried more transverse momentum than their containing jet. 
This excess was caused by corrections applied during jet reconstruction, which in some cases led to an underestimation of the jet energy,
and was well reproduced in simulations.

\begin {figure}
\includegraphics {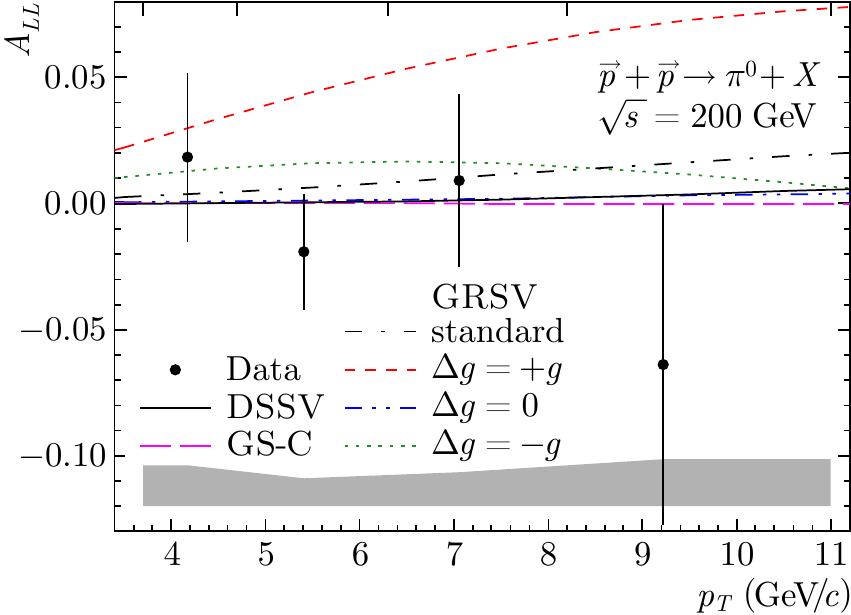}
\caption {\label {fig:All}(color online) Longitudinal double-spin asymmetry for inclusive $\pizero$ production
at midrapidity in \protonproton\ collisions at $\sqrts = 200\unit{\GeV}$,
compared to \NLO\ \pQCD\ calculations based on the gluon distributions 
from the \GRSV~\protect\cite {Gluck:2000dy}, \GSC~\protect\cite {Gehrmann:1995ag}, and \DSSV~\protect\cite {deFlorian:2008mr} global analyses. 
The systematic error (shaded band) does not include a $9.4\%$ normalization uncertainty due to the beam polarization measurement.}
\end {figure}

The asymmetry [Eq.~(\ref {all_def})] was calculated as
\begin {equation}
A_{LL} = \frac{1}{P_1 P_2}\frac{(N^{++} - RN^{+-})}{(N^{++} + RN^{+-})},
\end {equation}
where $N^{++}$ and $N^{+-}$ are the $\pizero$ yields in equal and opposite beam helicity configurations, respectively, 
and $R$ is the luminosity ratio for those two helicities. 
Typical values of $R$, measured with the \BBC{}s to a statistical precision of $10^{-3}$--$10^{-4}$ per run, ranged from $0.85$ to $1.2$, depending on fill and bunch pattern. 
Figure~\ref {fig:All} 
shows the measured longitudinal double-spin asymmetry for $\pizero$ production.
The data points are plotted at the mean pion $\pT$ in each bin. 
The lowest-$\pT$ point at $4.17\unit{\GeVc}$ was obtained from \HT1 triggers only; other points are the \HT1 and \HT2 combined results.

The systematic errors shown in the figure include point-to-point contributions from 
$\pizero$ yield extraction [(4--14)$\times 10^{-3}$], 
invariant mass background subtraction [(6--11)$\times 10^{-3}$], 
and remaining beam background [(1--9)$\times 10^{-3}$], 
as well as $\pT$-correlated contributions from 
relative luminosity uncertainties ($9\times 10^{-4}$) and from non-longitudinal spin components ($3\times 10^{-4}$).
All of the errors above are absolute errors on the measured asymmetry.
An evaluation of the effects of non-longitudinal components of the beam polarization was not possible 
due to the limited statistics of \pizero's in data taken with transversely polarized beams.
Instead, the largest value from the jet measurement~\cite {Abelev:2007vt} over the relevant momentum range was taken as an estimate of this systematic error. 
An overall normalization uncertainty of $9.4\%$ due to the uncertainty in the \RHIC\ \CNI\ polarimeter calibration is not shown. 
Studies of parity-violating single spin asymmetries and randomized spin patterns showed no evidence of bunch-to-bunch or fill-to-fill systematics. 
\clearpage

In Fig.~\ref {fig:All}, the measured values for $A_{LL}$ are compared to \NLO\ \pQCD\ calculations~\cite {Jager:2002xm} 
based on various sets of polarized gluon distribution functions. 
The \DSSV\ curve~\cite {deFlorian:2008mr} is the result of the first global analysis that includes semi-inclusive and inclusive \DIS\ data, as well as results obtained 
by the \PHENIX~\cite {Adare:2007dg} and \STAR~\cite {Abelev:2007vt} experiments. 
The \GSC\ curve~\cite {Gehrmann:1995ag} refers to a polarized gluon distribution function that has a large positive gluon polarization at low $x$, a node near $x \approx 0.1$, and a negative 
gluon polarization at large $x$\@.
The \GRSV\ standard curve is based on the best fit to \DIS\ data~\cite {Gluck:2000dy}, while the other \GRSV\ curves show scenarios of extreme positive ($\Delta g = +g$), 
extreme negative ($\Delta g = -g$), and vanishing ($\Delta g = 0$) gluon polarization at the starting scale~\cite {Gluck:2000dy,Jager:2004jh}\@.
A maximal positive gluon polarization scenario,
which has a total gluon spin contribution $\Delta G \equiv \int_{0}^{1}\! \Delta g(x)\, dx = 1.26$ 
at an initial scale of $0.4\unit{\GeV^2}$~\cite {Gluck:2000dy,Gluck:1998xa},
is excluded by our measurement at $98\unitns{\%}$ confidence level, including systematic uncertainties. 
This is in agreement with the conclusions from the inclusive jet measurements by \STAR~\cite {Abelev:2006uq,Abelev:2007vt} and from the inclusive \pizero\ measurement by \PHENIX~\cite {Adare:2007dg}\@. 
The data are consistent with all other gluon polarization scenarios, in particular with the \DSSV\ case.

In summary, we report a measurement of the invariant cross section and the longitudinal double-spin asymmetry $A_{LL}$ for inclusive 
\pizero\ production at midrapidity with the \STAR\ detector at \RHIC\@. The cross section was determined for $1 < \pT < 17\unit{\GeVc}$ 
and found to be in agreement with a \NLO\ \pQCD\ calculation based on the \CTEQsixM\ parton distribution functions and the \DSS\ fragmentation functions.
This set of fragmentation functions was constrained by data that included
measurements of \pizero\ production at midrapidity by \PHENIX~\cite {Adare:2007dg} and at forward rapidity
by \STAR~\cite {Adams:2006uz}\@.
The mean transverse momentum fraction of \pizero's in electromagnetically triggered jets was found to be approximately $0.7$ and to rise slightly with pion $\pT$,
in agreement with a \PYTHIA-based Monte Carlo simulation that included detector effects.
This measurement has the potential to contribute to future fragmentation function studies.
The asymmetry $A_{LL}$ was measured in the hard scattering regime at $3.7 < \pT < 11\unit{\GeVc}$ and found to be consistent with \NLO\ \pQCD\ calculations 
utilizing polarized quark and gluon distributions from inclusive and semi-inclusive \DIS\ data and from polarized proton data. 
Our data exclude a maximal positive gluon polarization in the nucleon, in agreement with results obtained from inclusive jet production in 
polarized proton collisions 
by \STAR~\cite {Abelev:2006uq, Abelev:2007vt}, 
while being a statistically independent measurement, subject to a different set of systematic uncertainties.
With increasing integrated luminosity, 
the neutral pion channel has the potential to provide additional constraints on the gluon polarization in the polarized proton.

We thank the \RHIC\ Operations Group and \RCF\ at \BNL, 
the \NERSC\ Center at \LBNL\ and the Open Science Grid consortium for providing resources and support. 
This work was supported in part by the Offices of \capsword{NP} and \capsword{HEP} within the U\tsp{-0.6}.\tsp{1}S.\ \capsword{DOE} Office of Science, 
the U\tsp{-0.6}.\tsp{1}S.\ \capsword{NSF}, the Sloan Foundation, the \capsword{DFG} cluster of excellence `Origin and Structure of the Universe', 
\capsword{CNRS/IN2P3}, \capsword{STFC} and \capsword{EPSRC} of the United Kingdom, \capsword{FAPESP} \capsword{CNP}q of Brazil, 
Ministry of Ed.\ and Sci.\ of the Russian Federation, 
\capsword{NNSFC}, \capsword{CAS}, \capsword{M}o\capsword{ST}, and \capsword{M}o\capsword{E} of China, 
\capsword{GA} and \capsword{MSMT} of the Czech Republic, 
\capsword{FOM} and \capsword{NWO} of the Netherlands, \capsword{DAE}, \capsword{DST}, and \capsword{CSIR} of India, 
Polish Ministry of Sci.\ and Higher Ed., 
Korea Research Foundation, Ministry of Sci., Ed.\ and Sports of the Rep.\ Of Croatia, 
Russian Ministry of Sci.\ and Tech, and RosAtom of Russia.


\begin{thebibliography}{35}
\expandafter\ifx\csname natexlab\endcsname\relax\def\natexlab#1{#1}\fi
\expandafter\ifx\csname bibnamefont\endcsname\relax
  \def\bibnamefont#1{#1}\fi
\expandafter\ifx\csname bibfnamefont\endcsname\relax
  \def\bibfnamefont#1{#1}\fi
\expandafter\ifx\csname citenamefont\endcsname\relax
  \def\citenamefont#1{#1}\fi
\expandafter\ifx\csname url\endcsname\relax
  \def\url#1{\texttt{#1}}\fi
\expandafter\ifx\csname urlprefix\endcsname\relax\fi
\providecommand{\bibinfo}[2]{#2}
\providecommand{\eprint}[2][]{\url{#2}}

\bibitem[{\citenamefont{Ashman et~al.}(1989)}]{Ashman:1989ig}
\bibinfo{author}{\bibfnamefont{J.}~\bibnamefont{Ashman}} \bibnamefont{et~al.}
  (\bibinfo{collaboration}{{EMC}}), \bibinfo{journal}{Nucl.
  Phys.} \textbf{\bibinfo{volume}{B328}}, \bibinfo{pages}{1}
  (\bibinfo{year}{1989}).

\bibitem[{\citenamefont{Filippone and Ji}(2001)}]{Filippone:2001ux}
\bibinfo{author}{\bibfnamefont{B.~W\!.}~\bibnamefont{Filippone}}~\bibnamefont{and}~\bibinfo{author}{\bibfnamefont{X.\mbox{-}D.}~\bibnamefont{Ji}},~\bibinfo{journal}{Adv.~Nucl.~Phys.}~\textbf{\bibinfo{volume}{26}},~\bibinfo{pages}{1}~(\bibinfo{year}{2001}).

\bibitem[{\citenamefont{Adeva et~al.}(1998)}]{Adeva:1998vw}
\bibinfo{author}{\bibfnamefont{B.}~\bibnamefont{Adeva}} \bibnamefont{et~al.}
  (\bibinfo{collaboration}{{SMC}}), \bibinfo{journal}{Phys.
  Rev.} \textbf{\bibinfo{volume}{D58}}, \bibinfo{pages}{112002}
  (\bibinfo{year}{1998}).

\bibitem[{\citenamefont{Anthony et~al.}(2000)}]{Anthony:2000fn}
\bibinfo{author}{\bibfnamefont{P.~L.} \bibnamefont{Anthony}}
  \bibnamefont{et~al.} (\bibinfo{collaboration}{{E155}}),
  \bibinfo{journal}{Phys. Lett.} \textbf{\bibinfo{volume}{B493}},
  \bibinfo{pages}{19} (\bibinfo{year}{2000}).

\bibitem[{\citenamefont{Airapetian et~al.}(2000)}]{Airapetian:1999ib}
\bibinfo{author}{\bibfnamefont{A.}~\!\bibnamefont{Airapetian}}~\bibnamefont{et~\tsp{-0.5}al.}~\!(\bibinfo{collaboration}{{HERMES}})\tsp{-0.5},~\!\bibinfo{journal}{Phys.~\!Rev\!.~\!Lett.}~\!\textbf{\bibinfo{volume}{84}},~\!\bibinfo{pages}{2584}~\!(\bibinfo{year}{2000}).

\bibitem[{\citenamefont{Adeva et~al.}(2004)}]{Adeva:2004dh}
\bibinfo{author}{\bibfnamefont{B.}~\bibnamefont{Adeva}} \bibnamefont{et~al.}
  (\bibinfo{collaboration}{{SMC}}), \bibinfo{journal}{Phys.
  Rev.} \textbf{\bibinfo{volume}{D70}}, \bibinfo{pages}{012002}
  (\bibinfo{year}{2004}).

\bibitem[{\citenamefont{Ageev et~al.}(2006)}]{Ageev:2005pq}
\mbox{\bibinfo{author}{\bibfnamefont{E.~\!S.}~\!\bibnamefont{Ageev}}~\!\bibnamefont{et~\!al.}~\!(\bibinfo{collaboration}{{COMPASS}}),~\!\bibinfo{journal}{Phys.~\!Lett.}~\!\textbf{\bibinfo{volume}{B633}},~\!\bibinfo{pages}{25}~\!(\bibinfo{year}{2006}).}

\bibitem[{\citenamefont{J{\"{a}}ger et~al.}(2003)\citenamefont{J{\"{a}}ger,
  Sch{\"{a}}fer, Stratmann, and Vogelsang}}]{Jager:2002xm}
\bibinfo{author}{\bibfnamefont{B.}~\bibnamefont{J{\"{a}}ger}},
  \bibinfo{author}{\bibfnamefont{A.}~\bibnamefont{Sch{\"{a}}fer}},
  \bibinfo{author}{\bibfnamefont{M.}~\bibnamefont{Stratmann}},
  \bibnamefont{and}
  \bibinfo{author}{\bibfnamefont{W\!.}~\bibnamefont{Vogelsang}},
  \bibinfo{journal}{Phys. Rev.} \textbf{\bibinfo{volume}{D67}},
  \bibinfo{pages}{054005} (\bibinfo{year}{2003}).

\bibitem[{\citenamefont{J{\"{a}}ger et~al.}(2004)\citenamefont{J{\"{a}}ger,
  Stratmann, and Vogelsang}}]{Jager:2004jh}
\bibinfo{author}{\bibfnamefont{B.}~\bibnamefont{J{\"{a}}ger}},
  \bibinfo{author}{\bibfnamefont{M.}~\bibnamefont{Stratmann}},
  \bibnamefont{and}
  \bibinfo{author}{\bibfnamefont{W\!.}~\bibnamefont{Vogelsang}},
  \bibinfo{journal}{Phys. Rev.} \textbf{\bibinfo{volume}{D70}},
  \bibinfo{pages}{034010} (\bibinfo{year}{2004}).

\bibitem[{\citenamefont{Adare et~al.}(2007)}]{Adare:2007dg}
\bibinfo{author}{\bibfnamefont{A.}~\bibnamefont{Adare}}~\bibnamefont{et~al.}~(\bibinfo{collaboration}{{PHENIX}}),~\bibinfo{journal}{Phys.~Rev.}~\textbf{\bibinfo{volume}{D76}},~\bibinfo{pages}{051106}~(\bibinfo{year}{2007}).

\bibitem[{\citenamefont{Abelev et~al.}(2008)}]{Abelev:2007vt}
\mbox{\bibinfo{author}{\bibfnamefont{B.~\!\!I.}~\!\bibnamefont{Abelev}}~\!\bibnamefont{et~\!al.}~\!(\bibinfo{collaboration}{{STAR}})\tsp{-0.3},~\!\bibinfo{journal}{Phys.~\!Rev\tsp{-0.5}.~\!Lett.}\tsp{0.5}\textbf{\bibinfo{volume}{100}}\tsp{-0.3},\tsp{0.5}\bibinfo{pages}{232003}~\tsp{-1.5}(\bibinfo{year}{2008}).}

\bibitem[{\citenamefont{de~Florian et~al.}(2008)\citenamefont{de~Florian,
  Sassot, Stratmann, and Vogelsang}}]{deFlorian:2008mr}
\bibinfo{author}{\bibfnamefont{D.}~\!\bibnamefont{de~\!Florian}},~\!\bibinfo{author}{\bibfnamefont{R.}~\!\bibnamefont{Sassot}},~\!\bibinfo{author}{\bibfnamefont{M.}~\!\bibnamefont{Stratmann}},~\!\bibnamefont{and}~\!\bibinfo{author}{\bibfnamefont{W\!.}~\!\bibnamefont{Vogelsang}},
  \bibinfo{journal}{Phys. Rev. Lett.} \textbf{\bibinfo{volume}{101}},
  \bibinfo{pages}{072001} (\bibinfo{year}{2008}).

\bibitem[{\citenamefont{Ackermann et~al.}(2003)}]{Ackermann:2002ad}
\mbox{\bibinfo{author}{\bibfnamefont{K.~H.}~\bibnamefont{Ackermann}}~\bibnamefont{et~al.}~(\bibinfo{collaboration}{{STAR}}),~\bibinfo{journal}{NIM}~\textbf{\bibinfo{volume}{A499}},~\bibinfo{pages}{624}~(\bibinfo{year}{2003}).}

\bibitem[{\citenamefont{Alekseev et~al.}(2003)}]{Alekseev:2003sk}
\bibinfo{author}{\bibfnamefont{I.}~\bibnamefont{Alekseev}}
  \bibnamefont{et~al.}, \bibinfo{journal}{NIM}
  \textbf{\bibinfo{volume}{A499}}, \bibinfo{pages}{392} (\bibinfo{year}{2003}).

\bibitem[{\citenamefont{Nakagawa et~al.}(2007)}]{Nakagawa:2007zza}
\bibinfo{author}{\bibfnamefont{I.}~\bibnamefont{Nakagawa}}
  \bibnamefont{et~al.}, \bibinfo{journal}{{AIP} {C}onf. Proc.}
  \textbf{\bibinfo{volume}{915}}, \bibinfo{pages}{912} (\bibinfo{year}{2007}).

\bibitem[{\citenamefont{Makdisi et~al.}(2007)}]{Makdisi:2007zz}
\bibinfo{author}{\bibfnamefont{Y\!.~I.} \bibnamefont{Makdisi}} \bibnamefont{et~al.}, \bibinfo{journal}{{AIP} {C}onf. Proc.}
  \textbf{\bibinfo{volume}{915}}, \bibinfo{pages}{975} (\bibinfo{year}{2007}).

\bibitem[{\citenamefont{Kiryluk}(2005)}]{Kiryluk:2005gg}
\bibinfo{author}{\bibfnamefont{J.}~\bibnamefont{Kiryluk}}
  (\bibinfo{collaboration}{{STAR}}),
  \bibinfo{journal}{{SPIN} 2004 Conf. Proc.}, \bibinfo{pages}{718}
  (\bibinfo{year}{2005}).

\bibitem[{\citenamefont{Beddo et~al.}(2003)}]{Beddo:2002zx}
\mbox{\bibinfo{author}{\bibfnamefont{M.}~\bibnamefont{Beddo}}~\bibnamefont{et~al.}~(\bibinfo{collaboration}{{STAR}}),~\bibinfo{journal}{NIM}~\textbf{\bibinfo{volume}{A499}},~\bibinfo{pages}{725}~(\bibinfo{year}{2003}).}

\bibitem[{\citenamefont{Anderson et~al.}(2003)}]{Anderson:2003ur}
\mbox{\bibinfo{author}{\bibfnamefont{M.}~\bibnamefont{Anderson}}~\bibnamefont{et~al.}~(\bibinfo{collaboration}{{STAR}}),~\bibinfo{journal}{NIM}~\textbf{\bibinfo{volume}{A499}},~\bibinfo{pages}{659}~(\bibinfo{year}{2003}).}

\bibitem[{\citenamefont{Adams et~al.}(2003)}]{Adams:2003kv}
\bibinfo{author}{\bibfnamefont{J.}~\bibnamefont{Adams}}~\bibnamefont{et~al.}~(\bibinfo{collaboration}{{STAR}}),~\bibinfo{journal}{Phys.~Rev.~Lett.}~\textbf{\bibinfo{volume}{91}},~\bibinfo{pages}{172302}~(\bibinfo{year}{2003}).

\bibitem[{\citenamefont{Grebenyuk}(2007)}]{ref_grebenyuk_thesis}
\bibinfo{author}{\bibfnamefont{O.}~\bibnamefont{Grebenyuk}}, Ph.D. thesis,
  \bibinfo{school}{{U}trecht {U}niversity} (\bibinfo{year}{2007}).

\bibitem[{\citenamefont{Abelev et~al.}()}]{pi0_pp_dau}
\bibinfo{author}{\bibfnamefont{B.~I.} \bibnamefont{Abelev}}
  \bibnamefont{et~al.} (\bibinfo{collaboration}{{STAR}}),
  \bibinfo{note}{to be published}.

\bibitem[{\citenamefont{Brun et~al.}(1978)\citenamefont{Brun, Hagelberg,
  Hansroul, and Lassalle}}]{Brun:1978fy}
\bibinfo{author}{\bibfnamefont{R.}~\bibnamefont{Brun}},
  \bibinfo{author}{\bibfnamefont{R.}~\bibnamefont{Hagelberg}},
  \bibinfo{author}{\bibfnamefont{M.}~\bibnamefont{Hansroul}}, \bibnamefont{and}
  \bibinfo{author}{\bibfnamefont{J.~C.} \bibnamefont{Lassalle}}
  (\bibinfo{year}{1978}), \bibinfo{note}{{CERN-DD-78-2-REV}}.

\bibitem[{\citenamefont{de~Florian et~al.}(2007)\citenamefont{de~Florian,
  Sassot, and Stratmann}}]{deFlorian:2007aj}
\bibinfo{author}{\bibfnamefont{D.}~\bibnamefont{de~Florian}},
  \bibinfo{author}{\bibfnamefont{R.}~\bibnamefont{Sassot}}, \bibnamefont{and}
  \bibinfo{author}{\bibfnamefont{M.}~\bibnamefont{Stratmann}},
  \bibinfo{journal}{Phys. Rev.} \textbf{\bibinfo{volume}{D75}},
  \bibinfo{pages}{114010} (\bibinfo{year}{2007}).

\bibitem[{\citenamefont{Adams et~al.}(2006{\natexlab{a}})}]{Adams:2006nd}
\bibinfo{author}{\bibfnamefont{J.}~\bibnamefont{Adams}} \bibnamefont{et~al.}
  (\bibinfo{collaboration}{{STAR}}), \bibinfo{journal}{Phys.
  Lett.} \textbf{\bibinfo{volume}{B637}}, \bibinfo{pages}{161}
  (\bibinfo{year}{2006}{\natexlab{a}}).

\bibitem[{\citenamefont{Pumplin et~al.}(2002)}]{Pumplin:2002vw}
\bibinfo{author}{\bibfnamefont{J.}~\bibnamefont{Pumplin}} \bibnamefont{et~al.},
  \bibinfo{journal}{JHEP} \textbf{\bibinfo{volume}{07}}, \bibinfo{pages}{012}
  (\bibinfo{year}{2002}).

\bibitem[{\citenamefont{Adams et~al.}(2006{\natexlab{b}})}]{Adams:2006uz}
\bibinfo{author}{\bibfnamefont{J.}~\bibnamefont{Adams}}~\bibnamefont{et~al.}~(\bibinfo{collaboration}{{STAR}}),~\bibinfo{journal}{Phys.~Rev.~Lett.}~\textbf{\bibinfo{volume}{97}},~\bibinfo{pages}{152302}~(\bibinfo{year}{2006}{\natexlab{b}}).

\bibitem[{\citenamefont{Abelev et~al.}(2006)}]{Abelev:2006uq}
\bibinfo{author}{\bibfnamefont{B.~I.}~\bibnamefont{Abelev}}~\bibnamefont{et~al.}~(\bibinfo{collaboration}{{STAR}}),~\bibinfo{journal}{Phys.~Rev.~Lett.}~\textbf{\bibinfo{volume}{97}},~\bibinfo{pages}{252001}~(\bibinfo{year}{2006}).

\bibitem[{\citenamefont{Boca et~al.}(1991)}]{Boca:1990rh}
\bibinfo{author}{\bibfnamefont{G.}~\bibnamefont{Boca}} \bibnamefont{et~al.},
  \bibinfo{journal}{Z. Phys.} \textbf{\bibinfo{volume}{C49}},
  \bibinfo{pages}{543} (\bibinfo{year}{1991}).

\bibitem[{\citenamefont{de~Florian}(2009)}]{deFlorian:2009fw}
\bibinfo{author}{\bibfnamefont{D.}~\bibnamefont{de~Florian}},
  \bibinfo{journal}{Phys. Rev.} \textbf{\bibinfo{volume}{D79}},
  \bibinfo{pages}{114014} (\bibinfo{year}{2009}).

\bibitem[{\citenamefont{Sj{\"{o}}strand
  et~al.}(2001)\citenamefont{Sj{\"{o}}strand, Ed{\'{e}}n, Friberg,
  L{\"{o}}nnblad, Miu, Mrenna, and Norrbin}}]{Sjostrand:2001yu}
\bibinfo{author}{\bibfnamefont{T\tsp{-0.5}.}~\tsp{-0.5}\bibnamefont{Sj{\"{o}}strand}},~\!\bibinfo{author}{\bibfnamefont{P\!.}~\!\bibnamefont{Ed{\'{e}}n}},~\!\bibinfo{author}{\bibfnamefont{C.}~\!\bibnamefont{Friberg}},~\!\bibinfo{author}{\bibfnamefont{L.}~\!\bibnamefont{L{\"{o}}nnblad}},~\!\bibinfo{author}{\bibfnamefont{G.}~\!\bibnamefont{Miu}},~\!\bibinfo{author}{\bibfnamefont{S.}~\!\bibnamefont{Mrenna}},
\bibnamefont{and}~\!\bibinfo{author}{\bibfnamefont{E.}~\!\bibnamefont{Norrbin}},~\!\bibinfo{journal}{Comput.~\!Phys.~\!Commun.}~\!\textbf{\bibinfo{volume}{135}},~\!\bibinfo{pages}{238}~\!(\bibinfo{year}{2001}).

\bibitem[{\citenamefont{Field and Group}(2006)}]{Field:2005sa}
\bibinfo{author}{\bibfnamefont{R.}~\bibnamefont{Field}} \bibnamefont{and}
  \bibinfo{author}{\bibfnamefont{R.~C.} \bibnamefont{Group}}
  (\bibinfo{collaboration}{{CDF}}),
  \eprint{hep-ph/0510198}.

\bibitem[{\citenamefont{Gl{\"{u}}ck et~al.}(2001)\citenamefont{Gl{\"{u}}ck,
  Reya, Stratmann, and Vogelsang}}]{Gluck:2000dy}
\bibinfo{author}{\bibfnamefont{M.}~\bibnamefont{Gl{\"{u}}ck}},
  \bibinfo{author}{\bibfnamefont{E.}~\bibnamefont{Reya}},
  \bibinfo{author}{\bibfnamefont{M.}~\bibnamefont{Stratmann}},
  \bibnamefont{and}
  \bibinfo{author}{\bibfnamefont{W\!.}~\bibnamefont{Vogelsang}},
  \bibinfo{journal}{Phys. Rev.} \textbf{\bibinfo{volume}{D63}},
  \bibinfo{pages}{094005} (\bibinfo{year}{2001}).

\bibitem[{\citenamefont{Gehrmann and Stirling}(1996)}]{Gehrmann:1995ag}
\mbox{\bibinfo{author}{\bibfnamefont{T.}~\!\bibnamefont{Gehrmann}}~\!\bibnamefont{and}~\!\bibinfo{author}{\bibfnamefont{W\!.~\!J.}~\!\bibnamefont{Stirling}},~\!\bibinfo{journal}{Phys.~\!Rev.}~\!\textbf{\bibinfo{volume}{D53}},~\!\bibinfo{pages}{6100}~\!(\bibinfo{year}{1996}).}

\bibitem[{\citenamefont{Gl{\"{u}}ck et~al.}(1998)\citenamefont{Gl{\"{u}}ck,
  Reya, and Vogt}}]{Gluck:1998xa}
\bibinfo{author}{\bibfnamefont{M.}~\bibnamefont{Gl{\"{u}}ck}},~\bibinfo{author}{\bibfnamefont{E.}~\bibnamefont{Reya}},~\bibnamefont{and}~\bibinfo{author}{\bibfnamefont{A.}~\bibnamefont{Vogt}},~\bibinfo{journal}{Eur.~Phys.~J.}~\textbf{\bibinfo{volume}{C5}},~\bibinfo{pages}{461}~(\bibinfo{year}{1998}).

\end{thebibliography}
\end {document}